\documentclass[letterpaper, 10 pt, conference]{ieeeconf}  % Comment this line out if you need a4paper

\IEEEoverridecommandlockouts                              % This command is only needed if 
\usepackage{graphics} % for pdf, bitmapped graphics files
\usepackage{epsfig} % for postscript graphics files
\usepackage{mathptmx} % assumes new font selection scheme installed
\usepackage{times} % assumes new font selection scheme installed
\usepackage{amsmath} % assumes amsmath package installed
\allowdisplaybreaks[4]
\usepackage{amssymb}  % assumes amsmath package installed
\usepackage{mathtools}
\usepackage{newtxtext}
\usepackage{newtxmath}
\usepackage{tabularx}
\usepackage[font=small]{caption}
\usepackage{multirow}
\usepackage{multicol}
\usepackage{booktabs}
\usepackage{comment}
\usepackage{cite}
\usepackage{url}
\usepackage{hyperref}
\usepackage{subcaption}
\usepackage{algorithm}
\usepackage{algpseudocode}
\usepackage{cleveref}
\usepackage[table]{xcolor}

\usepackage{caption}
\title{\LARGE \bf
Portfolio Design and Pricing for Multimodal \\ Mobility-on-Demand Services Considering Traveler Responses}

\author{Jungeun Lee$^{1}$, Jeong hwan Jeon$^{1, *}$ and Youngseo Kim$^{2, *}$% <-this % stops a space
\thanks{This research was supported by Korean Institute for Advancement of Technology (KIAT) grant funded by the Korea Government (MOTIE) (RS-2024-00435815, Human Resource Development Program for Industrial Innovation (Global)).}
\thanks{$^{1}$ Department of Electrical Engineering, Ulsan National Institute of
Science and Technology (UNIST), Ulsan 44919, Republic of Korea}%
\thanks{$^{2}$ Department of Civil and Environmental Engineering, University of California, Los Angeles (UCLA), CA 90095, USA}%
\thanks{* Corresponding authors: jhjeon@unist.ac.kr, youngseo@ucla.edu}%
}

\begin{document}

\maketitle
\thispagestyle{empty}
\pagestyle{empty}

% Abstract
\begin{abstract}
Mobility-on-demand (MoD) platforms can offer exclusive, pooled, and transit-connected services to accommodate different passenger needs. However, determining which services to offer and at what prices is challenging because profitability depends on both passenger choices and operational feasibility. This paper proposes a real-time framework that jointly optimizes service menus, fares, and vehicle trip plans for individual MoD requests. Passenger choices are modeled using a generalized nested logit (GNL) model that captures correlated substitution among service options based on vehicle sharing and transit use. To solve the resulting nonlinear fare-optimization problem, we use Fenchel duality to transform the nonlinear objective from the fare domain to the choice-probability domain, yielding a concave maximization involving a generalized entropy term. We evaluate the framework through operational simulations using real-world New York City trip records. Joint portfolio design and pricing improve operator service margins by $28.4\%$ and $8.5\%$ compared with optimizing only service menus and only fares, respectively. Furthermore, the proposed reformulation is up to $18.8$ times faster than solving the original pricing problem directly in fare space, improving computational efficiency for real-time MoD operations.
\end{abstract}

% Introduction
\section{INTRODUCTION}
Mobility-on-demand (MoD) platforms can offer a variety of service options to accommodate different passenger needs~\cite{wang2019ridesourcing}. In addition to exclusive door-to-door rides, operators may offer pooled door-to-door rides and transit-connected services that provide MoD rides to and from transit stations~\cite{stiglic2018enhancing, ma2019dynamic}. These options can improve the use of limited vehicle resources. Pooling allows multiple passengers to share a vehicle trip, while transit-connected services can reduce the portion of a journey served by MoD vehicles. These operational benefits are particularly valuable when vehicle availability is constrained by high demand or congestion. However, whether these operational benefits improve operator profitability depends on how passengers respond to the offered options. A lower-priced pooled or transit-connected service may attract passengers who would otherwise not use MoD, but it may also draw passengers away from higher-margin exclusive or door-to-door services. Thus, offering an additional option can affect overall MoD demand and how that demand is distributed across services.

Predicting how passengers choose among these service options is challenging because their decisions depend on both observable service attributes, such as fares, waiting times, in-vehicle travel times, detour times, and transfer times, and unobserved preferences, such as discomfort with vehicle sharing and concerns about transferring. Since the same unobserved preferences may influence multiple options that share a common feature, their random utility terms may be correlated. For example, passengers who dislike vehicle sharing may be less likely to choose either pooled door-to-door or pooled transit-connected rides, while those who dislike transferring may avoid both exclusive and pooled transit-connected rides. The commonly used multinomial logit (MNL) model assumes independent and identically distributed (i.i.d.) random utility terms and therefore fails to reflect similarities among service options when predicting passenger choice~\cite{train2009discrete}. Therefore, we adopt a generalized nested logit (GNL)~\cite{wen2001gnl} to model correlations among service options. This allows the operator to predict how the offered fares shift passenger demand toward other options that passengers perceive as closer substitutes.

However, which service options to offer and how to price them depend not only on passenger choices but also on operational conditions. Vehicle availability, traffic conditions, and transit schedules determine which trip plans are feasible for each service option. We therefore propose a real-time framework that jointly determines service menus, fares, and vehicle trip plans for individual MoD requests. For each request, the operator evaluates four service options: exclusive door-to-door, pooled door-to-door, exclusive transit-connected, and pooled transit-connected rides. Feasible trip plans are first generated for each option subject to passenger service constraints and operational conditions. Among the options with feasible plans, the operator jointly determines which options to offer, which trip plan to use for each offered option, and the corresponding fares to maximize expected service margins under the GNL choice model. Because fares affect both the margin earned from each service and the probabilities of choosing the offered options, the resulting pricing problem is nonlinear and computationally demanding. We therefore use Fenchel duality~\cite{fenchelduality} to reformulate the pricing problem in terms of choice probabilities and generalized entropy, enabling more efficient pricing optimization for real-time operations.

The main contributions of this paper are threefold.
\begin{enumerate}
    \item{\textbf{Joint optimization of service portfolio design and pricing.}} We develop a framework that jointly optimizes service menus, fares, and vehicle trip plans for each passenger to maximize expected service margins while satisfying operational constraints. Simulations using about $4{,}900$ real-world NYC requests demonstrate that the proposed optimization framework increases operator service margins by $28.4\%$ and $8.5\%$ compared with optimizing only service menus and only fares, respectively.
    \item{\textbf{Modeling passenger choices among multimodal service options.}} We model passenger choice using a GNL that captures correlations in unobserved preferences, such as discomfort with vehicle sharing and concerns about transit transfers. Unlike approaches that assume passengers accept the offered service~\cite{cdc1, cdc2}, our framework predicts passenger choices among the offered service options and an outside option. This allows service menu and pricing decisions to account for both service acceptance and how demand shifts among service options.
    \item{\textbf{Computationally efficient pricing through a generalized-entropy reformulation.}} We use Fenchel duality to reformulate the nonlinear pricing problem in terms of passenger choice probabilities using a convex generalized entropy function. The resulting objective is a concave maximization problem in the continuous probability variables, enabling computationally efficient pricing optimization for real-time decision-making.
\end{enumerate}

% Related Works
\section{RELATED WORKS}
Pricing under discrete choice models has been extensively studied in revenue management. Under the MNL model, expected profit is nonconcave in prices due to the bilinear interaction between prices and 
price-dependent choice probabilities but can become concave when reformulated in terms of choice probabilities~\cite{hanson1996optimizing}. Li and Huh~\cite{li2011pricing} extended the reformulation method to nested logit models. More generally, these reformulations can be interpreted through convex duality, which relates the consumer surplus function to a generalized entropy function of choice probabilities~\cite{fosgerau2020discrete}. This relationship makes it possible to replace prices with choice probabilities as optimization variables, and then recover the corresponding prices after optimization. Such reformulation approach has also been applied to transportation systems to improve computational tractability~\cite{kim25tcns}, but has not been extended to jointly optimize service portfolios and fares in real-time MoD operations.

In parallel, research on real-time MoD operations has focused on efficient vehicle matching and routing. Santi et al.~\cite{santi2014quantifying} proposed shareability networks to identify compatible passenger trips, while Alonso-Mora et al.~\cite{javier17tvgraph} decomposed the matching and routing problem into a feasible trip enumeration algorithm and integer linear programming for vehicle assignment. Subsequent studies have used approximate dynamic programming and reinforcement learning to mitigate myopic dispatching decisions by predicting future demand and vehicle availability~\cite{shah2020neural, qin2022reinforcement}. Other studies have extended MoD dispatching to transit-connected services that provide first-mile and last-mile rides~\cite{ma2019dynamic, stiglic2018enhancing, vakayil2017integrating}. However, these approaches primarily focus on vehicle assignment and routing without jointly optimizing service portfolios and fares under a passenger choice model.

Our work connects these two lines of research by integrating passenger-choice-based pricing with real-time MoD operations. This integration is challenging because service availability, travel times, and operating costs depend on vehicle assignment and feasible trip plans, while passenger choices depend on the fares and service attributes offered by the operator. We address these interactions by jointly optimizing service portfolios, fares, and trip plans. 
% The pricing problem is then reformulated in terms of choice probabilities and generalized entropy to improve computational tractability.

% Methodology
\section{METHODOLOGY}
\begin{figure}[t]
    \centering
    \includegraphics[width=0.48\textwidth]{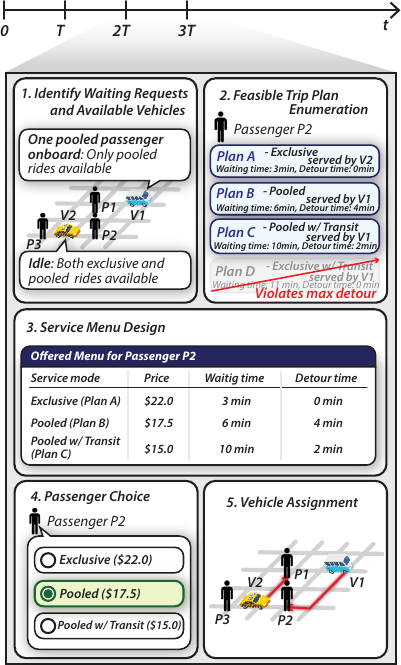}
    \caption{Overview of the proposed multimodal MoD portfolio and pricing framework. At each assignment time, the operator identifies requests awaiting service and available vehicles, generates feasible trip plans, and determines the service menu and fares for each passenger. Once passengers make their choices, vehicles are assigned to the selected services and begin serving them.}
    \label{fig:overview}
\end{figure}
As illustrated in Fig.~\ref{fig:overview}, the operator offers service menus at fixed intervals of length $T$, rather than immediately upon each request's arrival. This batch processing allows the operator to consider multiple requests together, reducing myopic vehicle allocation decisions that may arise when requests are processed individually upon arrival~\cite{javier17tvgraph}. At time $t_b=bT$, where $b$ is a positive integer, the operator identifies requests awaiting assignment, $\mathcal{R}_b$, and available vehicles, $\mathcal{V}_b$. The request set $\mathcal{R}_b$ includes requests received during $(t_{b-1},t_b]$ and previously unmatched requests that are still waiting for service. The available vehicles consist of idle vehicles with no existing passenger commitments, $\mathcal{V}_b^I$, and vehicles already assigned to pooled services with remaining capacity, $\mathcal{V}_b^P$. Thus, $\mathcal{V}_b=\mathcal{V}_b^I\cup\mathcal{V}_b^P$. Idle vehicles can provide either exclusive or pooled rides, whereas vehicles assigned to pooled services can accommodate additional passengers only if the resulting routes remain feasible for existing passengers.

The following subsections describe which service options the operator can offer, how feasible service trip plans are generated, and how passengers choose among the offered options. We then formulate an optimization problem to determine which service options to offer and at what prices, and propose a more computationally efficient reformulation of the original optimization problem.

\subsection{What service options can the operator offer?}
\begin{figure}[t]
    \centering
    \includegraphics[width=0.48\textwidth]{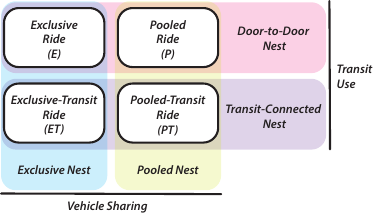}
    \caption{Generalized nested structure of the service options. Each service option belongs to two nests, one defined by vehicle sharing (exclusive or pooled) and the other by transit use (door-to-door or transit-connected).}
    \label{fig:cross_nested_structure}
\end{figure}
The operator provides mobility services on a road network where traffic congestion affects travel times. Each request $i$ specifies an origin $o_i$, a destination $d_i$, and a request time $t_i$. For each request $i$, the operator offers a subset of the four service options at the request time $t_i$. These options are exclusive door-to-door ($E$), pooled door-to-door ($P$), exclusive transit-connected ($ET$), and pooled transit-connected ($PT$) collectively denoted by $\mathcal{O} = \left\{E, P, ET, PT\right\}$. The four service options differ in vehicle sharing and transit use, as illustrated in Fig.~\ref{fig:cross_nested_structure}. Options $E$ and $P$ provide MoD service from the origin to the destination either using exclusive or pooled rides, respectively. For options $ET$ and $PT$, passengers take an MoD ride from their origin to the departure transit station, travel by transit to the arrival station, and then take another MoD ride to their destination. Options $E$ and $ET$ provide exclusive rides, whereas $P$ and $PT$ provide pooled rides.

\subsection{How are feasible trip plans generated for service options?}
To determine which service options can be offered to each passenger, the operator generates feasible trip plans for serving each request $i \in \mathcal{R}_b$ using the available vehicles in $\mathcal{V}_b$. A trip plan $\sigma_v$ for vehicle $v$ specifies the set of requests served, $\mathcal{K}\left(\sigma_v\right)$, and the ordered sequence of pickup and drop-off locations to be visited by the vehicle, denoted by $\boldsymbol{\pi}\left(\sigma_v\right)$.
For $E$, an idle vehicle serves request $i$ exclusively, with $\boldsymbol{\pi}\left(\sigma_v\right)=\left(o_i, d_i\right)$. For $P$, request $i$ may be assigned to an idle vehicle or added to the trip plan of a vehicle already assigned to a pooled service. Requests in the same batch are not pooled together since their service choices are still unknown, although a request may be pooled with requests from subsequent batches once its service choice is confirmed. Thus, each candidate pooled plan includes at most one request from the current batch. For example, if vehicle $v$ is already assigned to serve passenger $k$ from an earlier batch under pooled service, request $i$ from the current batch may be added to its trip plan, yielding $\boldsymbol{\pi}\left(\sigma_v\right)=\left(o_i, o_k, d_k, d_i\right)$, where the vehicle picks up $i$ before $k$ and drops off $k$ before $i$. Options $ET$ and $PT$ require two vehicle trip plans connected by a transit itinerary. Let $\sigma_{v^O}$ denote the first-mile trip plan from the origin to the departure station and $\sigma_{v^D}$ the last-mile trip plan from the arrival station to the destination, served by the first-mile feeder ($v^O$) and the last-mile feeder ($v^D$), respectively. Under $ET$, their stop sequences are $\boldsymbol{\pi}(\sigma_{v^O}) = (o_i, s^O)$ and $\boldsymbol{\pi}(\sigma_{v^D}) = (s^D, d_i)$, where $s^O$ and $s^D$ are the departure and arrival stations. Under $PT$, each plan allows passengers to share the vehicle with other passengers, for example through plans with $\boldsymbol{\pi}(\sigma_{v^O}) = (o_{i}, o_{k}, s^O)$ and $\boldsymbol{\pi}(\sigma_{v^D})=(s^D, d_{m}, d_{i})$, where passenger $i$ shares the ride to $s^O$ with $k$ and the ride from $s^D$ with $m$.

A service option can be offered only if its vehicle trip plans and its transit itinerary (for $ET$ and $PT$) satisfy the following conditions.
\begin{itemize}
    \item \textbf{Vehicle capacity.} The number of passengers on board vehicle $v$ does not exceed its capacity $C_v$ at any point along its trip plan.
    \item \textbf{Waiting and detour limits.} Each vehicle trip plan limits every passenger's waiting time for a MoD vehicle to $W$. For pooled rides, the in-vehicle time of each passenger does not exceed $(1 + \Delta)$ times the door-to-door travel time, where $\Delta$ is the maximum relative detour~\cite{bischoff2017city, zwick2021ride}. This relative detour-time limit prevents excessive travel delays on the shorter MoD legs. For $ET$ and $PT$, these limits apply separately to each MoD leg. These limits protect passengers from excessive travel detour when new requests are added.
    \item \textbf{Transfer feasibility.} For $ET$ and $PT$, the passenger must have enough time to complete all transfers. Specifically, the passenger must reach the departure platform before the transit service departs, have sufficient time for any intermediate transit connections, and reach the last-mile MoD pickup point before the scheduled pickup time.
    \item \textbf{Last-mile feeder commitment.} A vehicle scheduled to pick up a passenger after the transit trip may serve other passengers beforehand. However, these additional trips must not prevent the vehicle from reaching the scheduled pickup point on time.
\end{itemize}
For each request $i$ and service option $j \in \mathcal{O}$, we collect the plans satisfying these conditions in the feasible service-plan set $\Phi_i^j$. Each element $\phi \in \Phi_i^j$ specifies a feasible trip plan that can be executed if passenger $i$ selects option $j$. For $E$ and $P$, $\phi$ consists of a single vehicle trip plan $\sigma_v$. For $ET$ and $PT$, $\phi$ consists of a first-mile trip plan, a transit itinerary, and a last-mile trip plan, represented by $(\sigma_{v^O}^O, \sigma^T, \sigma_{v^D}^D)$, where $\sigma^T$ specifies the transit itinerary between the departure and arrival stations. Option $j$ can be offered only if $\Phi_i^j \neq \varnothing$.

\subsection{How are passenger choices modeled?}
\label{sec:choice_model}
Passenger $i$ chooses among the service options included in the offered menu and an outside transit option, such as bus or subway. This specification focuses on public transit as the primary alternative to the considered MoD and transit-connected services. Consider a menu $\mathcal{S}_i$, a feasible service plan $\phi_i^j \in \Phi_i^j$ for each $j \in \mathcal{S}_i$, and the corresponding MoD fares $\boldsymbol{f}_i = (f_i^j)_{j \in \mathcal{S}_i}$. Each fare $f_i^j$ is determined conditional on the designated service plan $\phi_i^j$, although we omit this dependence from the notation for simplicity. We use a random-utility model~\cite{ben1985discrete} to predict which option the passenger will choose. The utility of each offered option $j \in \mathcal{S}_i$ is 
\begin{equation}
    U_i^j = V_i^j + \varepsilon_i^j,
\end{equation}
where $V_i^j$ is the deterministic utility and $\varepsilon_i^j$ represents factors unobserved by the operator, such as discomfort with vehicle sharing and concerns about transferring. To distinguish the MoD fare determined by the operator from the remaining service attributes, the deterministic utility is expressed as $V_i^j = a_i^j (\phi_i^j) - \alpha_f f_i^j$,
where $f_i^j$ is the MoD fare determined by the operator, while $a_i^j(\phi_i^j)$ collects the remaining utility components associated with option $j$ and its designated service plan $\phi_i^j$. Specifically, 
% $a_i^j (\phi_i^j) = \text{ASC}^j -\alpha_f f_i^{j,t}-\alpha_w w_i^j (\phi_i^j)-\alpha_v v_i^j (\phi_i^j) -\alpha_\delta \delta_i^j (\phi_i^j) -\alpha_t t_i^j (\phi_i^j),$
\begin{equation}
    \begin{aligned}
        a_i^j (\phi_i^j)
        ={}& \text{ASC}^j
        -\alpha_f f_i^{j,t}
        -\alpha_w w_i^j (\phi_i^j)
        -\alpha_v v_i^j (\phi_i^j) \\
        &-\alpha_\delta \delta_i^j (\phi_i^j)
        -\alpha_t t_i^j (\phi_i^j).
    \end{aligned}
    \label{eq:utility}
\end{equation}
Here, $\text{ASC}^j$ is the alternative-specific constant (ASC), and $f_i^{j,t}$, $w_i^j(\phi_i^j)$, $v_i^j(\phi_i^j)$, $\delta_i^j(\phi_i^j)$, and $t_i^j(\phi_i^j)$ denote the transit fare, waiting time, in-vehicle time, detour time, and transfer time associated with option $j$, respectively. The transit fare $f_i^{j,t}$ is zero for $E$ and $P$. For $ET$ and $PT$, it is paid to the transit agency and therefore affects passenger utility but is not included in the MoD operator's revenue. The positive coefficients $\alpha_f$, $\alpha_w$, $\alpha_v$, $\alpha_\delta$, and $\alpha_t$ represent sensitivities to these attributes.
Among these attributes, waiting and detour times are treated differently for exclusive and pooled options. For exclusive options ($E$ and $ET$), we use the estimated waiting times of the designated trip plans and set detour times to zero. For pooled options ($P$ and $PT$), the vehicle route may be updated after service selection as additional passengers are assigned, which can increase the passenger's waiting and detour times. We therefore model passengers as evaluating pooled options using the maximum waiting and detour times guaranteed by the operator rather than the values estimated from the initially designated plans. These guarantees must be satisfied whenever the trip plans are subsequently updated. The outside option is a public transit trip without MoD service. Its deterministic utility is given by $V_i^o = \text{ASC}^o - \alpha_f f_i^o - \alpha_w w_i^o - \alpha_v v_i^o - \alpha_\omega \omega_i^o - \alpha_t t_i^o$, where $f_i^o$, $w_i^o$, $v_i^o$, and $t_i^o$ denote the transit fare, waiting time, in-vehicle time, and transfer time, and $\omega_i^o$ denotes walking time. The coefficient $\alpha_\omega>0$ represents sensitivity to walking time.

If the random utility terms $\varepsilon_i^j$ are independent and identically distributed (i.i.d.) with type-I extreme-value distributions, the resulting choice model is multinomial logit (MNL). Under MNL, when the fare of one option increases, passengers switch to other options in proportion to their existing choice probabilities, regardless of how similar the services are. For example, if $P$ and $ET$ initially have equal shares, an increase in the fare of $E$ shifts equal amounts of demand to $P$ and $ET$ under the MNL model, even though unobserved discomfort with vehicle sharing may make $E$ and $ET$ closer substitutes, while concerns about transferring may create similar substitution patterns between $E$ and $P$. These shared attributes can induce correlations among the random utility terms. Therefore, we use a generalized nested logit model (GNL)~\cite{wen2001gnl} to capture these correlations. As shown in Fig.~\ref{fig:cross_nested_structure}, the four service options are grouped along two dimensions. The vehicle-sharing dimension groups the options into an exclusive nest $\mathcal{N}_E = \{E, ET\}$ and a pooled nest $\mathcal{N}_P = \{P, PT\}$, and the transit-use dimension groups them into a door-to-door nest $\mathcal{N}_D = \{E, P\}$ and a transit-connected nest $\mathcal{N}_T = \{ET, PT\}$. We denote the collection of these four nests by $\mathcal{N} = \{\mathcal{N}_E, \mathcal{N}_P, \mathcal{N}_D, \mathcal{N}_T\}$. Each option belongs to one nest in each dimension. Let $\gamma_{jn} \in [0, 1]$ denote the allocation weight of option $j$ to nest $n$, with $\gamma_{jn} = 0$ if $j \notin n$. For each option, the allocation weights across its nests sum to one. Each nest $n$ also has a dissimilarity parameter $\mu_n \in (0, 1]$, which controls the dependence among the random utility terms of options belonging to that nest. Smaller values of $\mu_n$ allow stronger dependence through the nest.

For a given menu $\mathcal{S}_i$, fares $\boldsymbol{f}_i$, and designated service plans $\boldsymbol{\phi}_i$, the choice probabilities under the GNL model are given by
\begin{equation}
p_i^j \left(\boldsymbol{f}_i, \mathcal{S}_i, \boldsymbol{\phi}_i\right) = \sum_{\substack{n\in\mathcal{N}:~j \in n \\ \gamma_{jn}>0}} P_i(n)P_i(j\mid n), \quad \forall~j \in \mathcal{S}_i,
\end{equation}
where $P_i(n)$ is the probability assigned to nest $n$, and $P_i(j \mid n)$ is the conditional probability of choosing option $j$ within that nest. These probabilities are given by
\begin{equation}
    P_i(n) = \frac{\left[\sum_{k \in n \cap \mathcal{S}_i} \left(\gamma_{kn} \exp{\left(V_i^k\right)}\right)^{1/\mu_n}\right]^{\mu_n}}{\exp{\left(V_i^o\right)} + \sum_{m \in \mathcal{N}} \left[\sum_{k \in m \cap \mathcal{S}_i} \left(\gamma_{km}\exp{\left(V_i^k\right)}\right)^{1/\mu_m}\right]^{\mu_m}},
\end{equation}
and
\begin{equation}
    P_i(j \mid n) = \frac{\left(\gamma_{jn}\exp{\left(V_i^j\right)}\right)^{1/\mu_n}}{\sum_{k \in n \cap \mathcal{S}_i} \left(\gamma_{kn}\exp{\left(V_i^k\right)}\right)^{1/\mu_n}}.
\end{equation}
The outside option is chosen with probability
\begin{equation}
    p_i^o = \frac{\exp{\left(V_i^o\right)}}{\exp{\left(V_i^o\right) + \sum_{n \in \mathcal{N}} \left[\sum_{j \in n \cap \mathcal{S}_i} \left(\gamma_{jn}\exp{\left(V_i^j\right)}\right)^{1/\mu_n}\right]^{\mu_n}}}
\end{equation}
For notational convenience, we define the exponentiated inclusive value for nest $n$ as
\begin{equation}
    I_{in} = \left[\sum_{j \in n \cap \mathcal{S}_i} \left(\gamma_{jn} \exp{\left(V_i^j\right)}\right)^{1/\mu_n}\right]^{\mu_n}, \quad \forall~n \in \mathcal{N}.
    \label{eq:inclusive}
\end{equation}
Only options included in $\mathcal{S}_i$ contribute to $I_{in}$. Thus, $I_{in} = 0$ if $n \cap \mathcal{S}_i = \varnothing$ or $\gamma_{jn} = 0$ for all $j \in n$. Using $I_{in}$, the nest and outside-option probabilities can be written more compactly as 
\begin{equation}
    P_i(n) = \frac{I_{in}}{\exp{\left(V_i^o\right) + \sum_{m \in \mathcal{N}} I_{im}}}, \quad p_i^o = \frac{\exp{\left(V_i^o\right)}}{\exp{\left(V_i^o\right) + \sum_{m \in \mathcal{N}} I_{im}}},
\end{equation}
which is subsequently used to define the GNL surplus function and reformulate the pricing problem in terms of choice probabilities.

\subsection{Which service options should the operator offer, and at what prices?}
Given the feasible service plans and passenger choice model defined above, the operator jointly determines the service menus, designated trip plans, and corresponding MoD fares for the requests in each batch. Let $f_i^j \geq 0$ denote the MoD fare charged to passenger $i$ for service option $j$. These decisions maximize the total expected service margin, which depends on the choice probability and incremental operating cost of each offered option. At time $t_b$, the incremental operating cost of serving request $i$ through plan $\phi\in\Phi_i^j$ under option $j$ is $c_{i,b}^j(\phi)=c_b(\phi)-c_b(\phi^{-i})$, where $\phi^{-i}$ denotes the same vehicle plan with request $i$ removed, and $c_b(\cdot)$ denotes the operating cost of the MoD legs, including labor, fuel, and vehicle depreciation. For pooled services, this difference excludes the cost that would still be incurred to serve previously accepted passengers. For transit-connected services, it includes the incremental costs of both the first-mile and last-mile MoD legs.

For each request $i\in\mathcal R_b$, let $y_i^j \in  \{0,1\}$ indicate whether service option $j$ is offered. The resulting menu is $\mathcal S_i(\boldsymbol y_i) = \{j\in\mathcal{O}: y_i^j=1\}$. With four service options, there are up to $2^{|\mathcal O|}=16$ possible menus, including the empty menu. For each offered option $j$, the operator designates exactly one feasible plan $\phi\in\Phi_i^j$. Let $x_{i\phi}^j \in \{0,1\}$ indicate whether plan $\phi$ is designated for request $i$ under option $j$. Since the passenger selects at most one offered option, the same vehicle can appear in the designated plans of multiple options for the same request. However, allowing the same vehicle to appear in plans offered to different requests could create a conflict if those passengers select those options simultaneously. To prevent such conflicts, let $A_{\phi v}$ indicate whether plan $\phi$ uses vehicle $v$. For $ET$ and $PT$, $A_{\phi v} = 1$ if vehicle $v$ serves either the first-mile or the last-mile MoD leg. We further define $u_{iv} \in \{0,1\}$ to indicate whether vehicle $v$ appears in any plan designated for request $i$. The vehicle-assignment constraints introduced below use these variables to ensure that each vehicle is associated with at most one request within a batch. For each offered option $j$, the designated plan determines the incremental operating cost incurred if the passenger selects that option: $c_{i,b}^j(\boldsymbol{x}_i) = \sum_{\phi \in \Phi_i^j} c_{i,b}^j(\phi)x_{i\phi}^j$.
The operator earns the fare and incurs the operating cost of option $j$ only when passenger $i$ selects that option. Therefore, its expected service margin is weighted by the corresponding choice probability. Since the offered menu and designated plans are determined by $\boldsymbol{y}_i = (y_i^j)_{j \in \mathcal{O}}$ and $\boldsymbol{x}_i = (x_{i\phi}^j)_{j \in \mathcal{O}, \phi \in \Phi_i^j}$, respectively, we write this probability as $p_i^j(\boldsymbol{f}_i,\boldsymbol{y}_i,\boldsymbol{x}_i)$. For brevity, we omit the arguments of $p_i^j$ in the following formulation.

Given these definitions, the operator solves the following optimization problem in each batch:
\begin{subequations}
\label{eq:ours}
\begin{align}
\max_{\boldsymbol f,\boldsymbol y,\boldsymbol x,\boldsymbol u}~& \sum_{i\in\mathcal R_b} \sum_{j\in\mathcal O} p_i^j \left[f_i^j-c_{i,b}^j(\boldsymbol x_i)\right] \label{eq:menu_obj} \\
\text{s.t.}~& \sum_{\phi\in\Phi_i^j} x_{i\phi}^j = y_i^j, && \forall i\in\mathcal R_b,~j\in\mathcal O, \label{eq:menu_plan} \\
& p_i^j \geq \underline{p} y_i^j, && \forall i \in \mathcal{R}_b,~j\in\mathcal{O}, \label{eq:menu_relevance} \\
& A_{\phi v}x_{i\phi}^j \leq u_{iv}, && \begin{gathered} \forall i\in\mathcal R_b,~j\in\mathcal O,\\ \phi\in\Phi_i^j,~v\in\mathcal V_b,\end{gathered} \label{eq:vehicle_link_lower} \\
& u_{iv} \leq \sum_{j\in\mathcal{O}} \sum_{\phi\in\Phi_i^j} A_{\phi v}x_{i\phi}^j, && \forall i\in\mathcal{R}_b,~v\in\mathcal{V}_b, \label{eq:vehicle_link_upper} \\
& \sum_{i\in\mathcal{R}_b} u_{iv} \leq 1,
&& \forall v\in\mathcal{V}_b, \label{eq:vehicle_assignment} \\
& f_i^j \geq 0, && \forall i\in\mathcal{R}_b,~j\in\mathcal{O}, \label{eq:fare_nonnegative} \\
& y_i^j\in\{0,1\}, && \forall i\in\mathcal{R}_b,~j\in\mathcal{O}, \label{eq:menu_binary} \\
& x_{i\phi}^j\in\{0,1\}, && \begin{gathered} \forall i\in\mathcal{R}_b,~j\in\mathcal{O}, \\ \phi\in\Phi_i^j,\end{gathered} \label{eq:plan_binary} \\
& u_{iv}\in\{0,1\}, && \forall i\in\mathcal{R}_b,~v\in\mathcal{V}_b. \label{eq:vehicle_binary}
\end{align}
\end{subequations}
Objective~\eqref{eq:menu_obj} maximizes the total expected service margin, where the margin of each option is weighted by its predicted choice probability. Constraint~\eqref{eq:menu_plan} requires exactly one feasible plan for each offered option and excludes options without feasible plans. If no service option is feasible, the passenger receives an empty menu and chooses the outside option, yielding zero service margin to the MoD operator.
Constraint~\eqref{eq:menu_relevance} requires each offered option to have a predicted choice probability of at least $\underline{p}>0$. This prevents the operator from including an option in the menu while setting unreasonably high fares for offered options. Constraints~\eqref{eq:vehicle_link_lower} and~\eqref{eq:vehicle_link_upper} ensure that the vehicle-inclusion indicator $u_{iv}$ equals $1$ if any plan designated for request $i$ uses vehicle $v$, and $u_{iv}=0$, otherwise.
Constraint~\eqref{eq:vehicle_assignment} then restricts each vehicle to at most one request in the current batch.
Constraint~\eqref{eq:fare_nonnegative} requires nonnegative MoD fares, and constraints~\eqref{eq:menu_binary}--\eqref{eq:vehicle_binary} define the binary decision variables.

\subsection{How can the pricing problem be reformulated for efficient optimization?}
The main computational challenge in problem~\eqref{eq:ours} arises from its nonlinear objective function. The objective maximizes the expected service margin, where fares enter in two ways: directly through the margin earned from each selected service and indirectly through the passenger choice probabilities. While a higher fare increases the margin earned from each selected service, it also changes the probabilities of choosing that option and the other offered options through the GNL choice model. This introduces products of fares and fare-dependent choice probabilities, creating the nonlinearity. To address this difficulty, we use Fenchel duality~\cite{fenchelduality} to reformulate the expected service margin in terms of choice probabilities, thereby removing the fare–probability product.

We begin with the GNL surplus function, $W_i(\boldsymbol{V}_i) = \ln\left[\exp{\left(V_i^o\right)} + \sum_{n\in\mathcal N}I_{in}\right]$, where $\boldsymbol{V}_i=(V_i^j)_{j\in\mathcal S_i}$ is the vector of deterministic utilities of the offered options, and $I_{in}$ is defined in Eq.~\eqref{eq:inclusive}. The choice probabilities satisfy $p_i^j = \partial W_i / \partial V_i^j$, so the vector $\boldsymbol{p}_i = (p_i^j)_{j \in \mathcal{S}_i}$ is the gradient of $W_i$ with respect to $\boldsymbol{V}_i$. The relationship motivates using $\boldsymbol{p}_i$ as the dual variable of $\boldsymbol{V}_i$ and introducing the convex conjugate of $W_i$, 
\begin{equation}
    W_i^*(\boldsymbol{p}_i) = \sup_{\boldsymbol{V}_i} \left\{ \sum_{j\in\mathcal{S}_i}p_i^jV_i^j - W_i(\boldsymbol{V}_i) \right\}.
\end{equation}
For the probabilities generated by the GNL model, the Fenchel--Young equality~\cite{fenchelduality} gives 
\begin{equation}
    \sum_{j \in \mathcal{S}_i} p_i^j V_i^j = W_i^*(\boldsymbol{p}_i) + W_i(\boldsymbol{V}_i).
\end{equation}
The outside-option probability is $p_i^o = \exp{\left(V_i^o - W_i(\boldsymbol{V}_i)\right)}$, which implies $W_i(\boldsymbol{V}_i) = V_i^o - \ln{p_i^o}$. Therefore,
\begin{equation}
    \sum_{j\in\mathcal{S}_i}p_i^jV_i^j = W_i^*(\boldsymbol{p}_i) + V_i^o -\ln p_i^o.
    \label{eq:fenchel_equality}
\end{equation}
For a given menu and designated plans, the contribution of passenger $i$ to the objective in Eq.~\eqref{eq:menu_obj} is $\sum_{j \in \mathcal{S}_i} p_i^j [f_i^j - c_{i,b}^j (\phi_i^j)]$, where $\phi_i^j$ denotes the plan designated for option $j$. Options not included in $\mathcal{S}_i$ have zero choice probability and therefore do not contribute to the objective.
Substituting $f_i^j = (a_i^j(\phi_i^j) - V_i^j) / \alpha_f$ into the expected service margin and applying Eq.~\eqref{eq:fenchel_equality} allows us to eliminate the fares:
\begin{equation}
\begin{aligned}
    &\sum_{j\in\mathcal{S}_i} p_i^j \left[ f_i^j-c_{i,b}^j(\phi_i^j) \right] \\
    & = \frac{1}{\alpha_f} \left[ \sum_{j\in\mathcal{S}_i} p_i^j \left[a_i^j(\phi_i^j) - \alpha_f c_{i,b}^j(\phi_i^j)\right] +\ln p_i^o - W_i^*(\boldsymbol p_i) - V_i^o
    \right].
\end{aligned}
\label{eq:fenchel_margin}
\end{equation}
The expected service margin is therefore expressed using the choice probabilities, the service attributes, and the incremental operating costs, without fare variables in the objective. 

To express $W_i^*(\boldsymbol{p}_i)$ in terms of generalized entropy, we introduce auxiliary variables. Let $\mathcal{A}=\{(j,n): j\in\mathcal{O}, j\in n, \gamma_{jn}>0\}$ denote the set of active option--nest pairs with positive allocation weights. For each $(j,n) \in \mathcal{A}$, let $z_{ijn}$ denote the probability that passenger $i$ chooses option $j$ through nest $n$. The choice probability of option $j$ is then $p_i^j = \sum_{\substack{n: (j, n) \in \mathcal{A}}} z_{ijn}$, and the total probability assigned to nest $n$ is $r_{in} = \sum_{j: (j,n) \in \mathcal{A}}z_{ijn}$. For an option that is not included in the offered menu, the constraint $p_i^j\le y_i^j$ forces $p_i^j=0$, which in turn forces $z_{ijn}=0$ for all nests containing that option.
The convex conjugate admits the representation
\begin{equation}
    W_i^*(\boldsymbol p_i) = \min_{\boldsymbol z_i} \left\{\Psi_i(\boldsymbol z_i,p_i^o) - p_i^oV_i^o\right\},
    \label{eq:gnl_conjugate}
\end{equation}
where
\begin{equation}
\begin{aligned}
    \Psi_i(\boldsymbol{z}_i,p_i^o) = & p_i^o\ln p_i^o + \sum_{n\in\mathcal{N}} \Bigg[(1-\mu_n)r_{in}\ln r_{in} \\
    & + \mu_n \sum_{j: (j, n) \in \mathcal{A}} z_{ijn} \ln z_{ijn} - \sum_{j: (j, n) \in \mathcal{A}} z_{ijn}\ln\gamma_{jn} \Bigg],
\end{aligned}
\label{eq:generalized_entropy}
\end{equation}
subject to $z_{ijn} \geq 0$, $p_i^o > 0$, and $p_i^o + \sum_{j \in \mathcal{O}} p_i^j = 1$, with convention $x \ln x = 0$ at $x = 0$. To represent the choice probability associated with each designated plan, we introduce the continuous variable $q_{i\phi}^j = p_i^j x_{i\phi}^j$ for $\phi \in \Phi_i^j$.
Since exactly one candidate plan is designated for each offered option, this product can be represented by $\sum_{\phi\in\Phi_i^j} q_{i\phi}^j=p_i^j$ and $0 \leq q_{i\phi}^j \leq x_{i\phi}^j$.
Thus, $q_{i\phi}^j$ equals the choice probability of option $j$ when plan $\phi$ is designated, and zero otherwise. Using Eq.~\eqref{eq:gnl_conjugate}, we obtain the following reformulated objective:
\begin{equation}
\begin{aligned}
    \max\quad \frac{1}{\alpha_f} \sum_{i\in\mathcal{R}_b} \Bigg[&\sum_{j\in\mathcal{O}} \sum_{\phi\in\Phi_i^j} q_{i\phi}^j \Big[a_i^j(\phi) - V_i^o - \alpha_f c_{i,b}^j(\phi)\Big] \\
    & + \ln p_i^o - \Psi_i(\boldsymbol{z}_i,p_i^o) \Bigg].
\end{aligned}
\label{eq:reformulated_objective}
\end{equation}
The reformulated problem retains constraints~\eqref{eq:menu_plan} and~\eqref{eq:vehicle_link_lower}--\eqref{eq:vehicle_binary}. Constraint~\eqref{eq:menu_relevance} is replaced by  $\underline{p}y_i^j \leq p_i^j\leq y_i^j$ for all $i\in\mathcal{R}_b$ and $j\in\mathcal{O}$, which ensures that an option has positive choice probability only if it is included in the offered menu.
Since $\Psi_i (\boldsymbol{z}_i, p_i^o)$ is jointly convex in $\boldsymbol{z}_i$ and $p_i^o$, the reformulated objective is concave in the continuous probability variables $q_{i\phi}^j$, $z_{ijn}$, and $p_i^o$. The menu, plan, and vehicle decisions $y_i^j$, $x_{i\phi}^j$, and $u_{iv}$ remain binary. To preserve equivalence with the original problem, we therefore impose a constraint that ensures the fares recovered from the optimal choice probabilities remain nonnegative. Let $\boldsymbol{p}_i^*$ and $\boldsymbol{x}_i^*$ denote the optimal choice probabilities and designated-plan decisions obtained from the reformulated problem. 
By conjugate duality, the deterministic utility corresponding to $\boldsymbol{p}_i^*$ satisfies $V_i^j = \partial W_i^*(\boldsymbol{p}_i^*) / \partial p_i^j$. Substituting this relation into
\begin{equation}
    V_i^j = \sum_{\phi \in \Phi_i^j} a_i^j(\phi) x_{i\phi}^{j,*} - \alpha_f f_i^j 
\end{equation}
gives the corresponding fare for each offered option $j$ as
\begin{equation}
    f_i^{j,*} = \frac{\sum_{\phi \in \Phi_i^j} a_i^j(\phi)x_{i\phi}^{j,*} - \dfrac{\partial W_i^*(\boldsymbol{p}_i^*)}{\partial p_i^j}}{\alpha_f}.
    \label{eq:fare_recovery}
\end{equation}
To ensure $f_i^j \geq 0$, the reformulated problem imposes
\begin{equation}
    \frac{\partial W_i^*(\boldsymbol{p}_i)}{\partial p_i^j} \leq \sum_{\phi \in \Phi_i^j} a_i^j(\phi) x_{i\phi}^j
\end{equation}
for every offered option $j$.

% Experimental results
\section{EXPERIMENTAL RESULTS}
\subsection{Experimental setup}
\begin{figure}[t]
    \centering
    \includegraphics[width=0.48\textwidth]{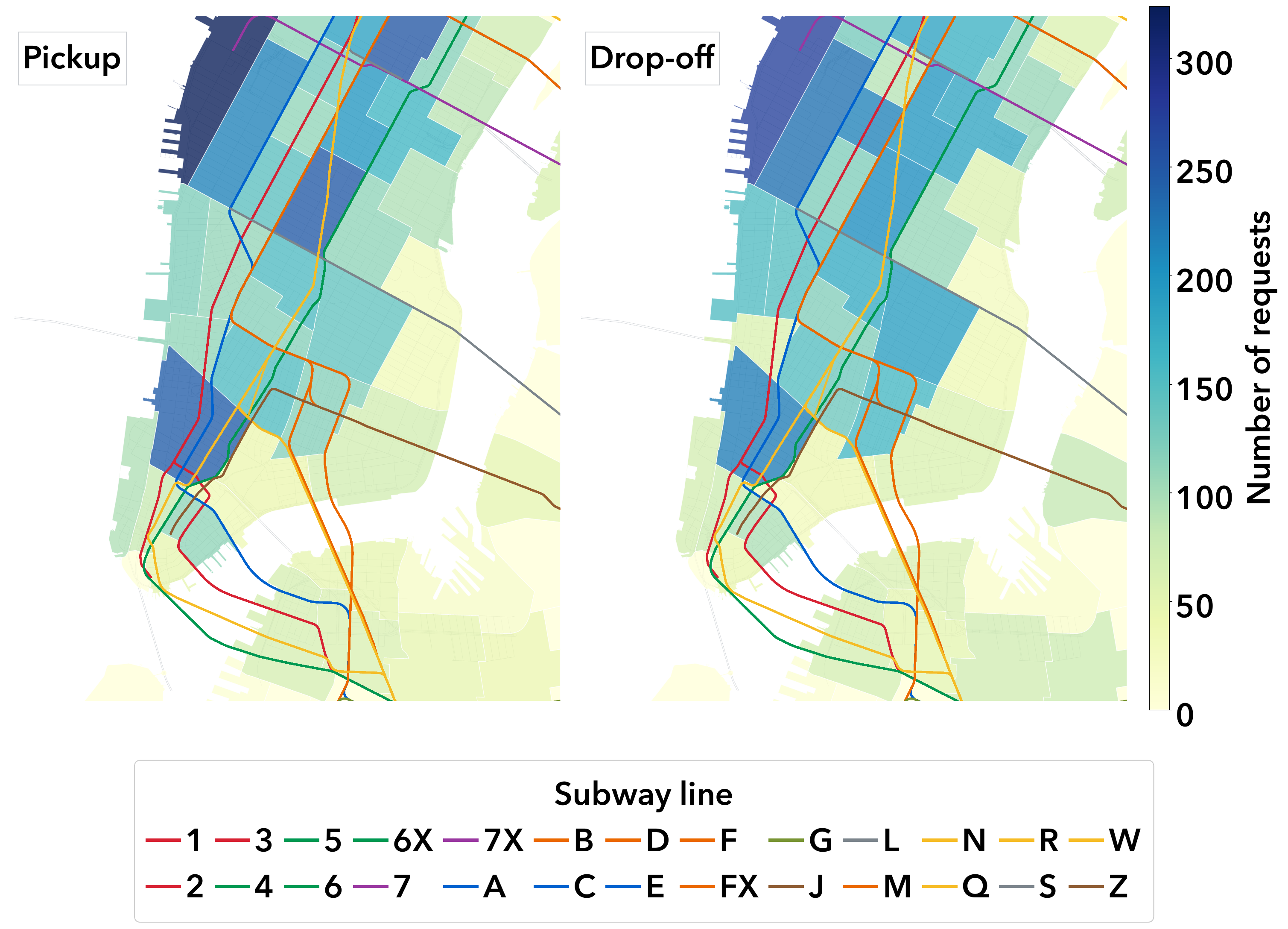}
    \caption{Study area and spatial distributions of travel demand. Pickup and drop-off demand during 4:00--5:00 PM on September 11, 2024, obtained from NYC TLC high-volume for-hire vehicle records~\cite{nyctlc}, are aggregated by taxi zone. NYC subway lines are overlaid on the maps.}
    \label{fig:od_dist}
\end{figure}

We simulate one hour of MoD operations in Manhattan, New York City (NYC). Passenger demand is obtained from the NYC Taxi and Limousine Commission (TLC) high-volume for-hire vehicle records~\cite{nyctlc} for Wednesday, September 11, 2024, between 4:00 PM and 5:00 PM. In the publicly released TLC data, pickup and drop-off locations are reported by taxi zone rather than exact coordinates. We retain all $4{,}900$ trips with both origin and destination zones within the study area, whose spatial distributions are shown in Fig.~\ref{fig:od_dist}. Vehicle routes are calculated using the OpenStreetMap driving network~\cite{osm}, while transit-connected services are constructed using transit lines and schedules from the Metropolitan Transportation Authority (MTA)~\cite{mtagtfs}. Based on an average request-to-vehicle ratio of approximately 11:1 estimated from NYC TLC data~\cite{nyctlc}, we set the fleet size to $430$ available MoD vehicles.

The batch interval is $ T=30$ s, the maximum waiting time is $W=10$ min, and the maximum detour ratio is $\Delta=0.5$. We set the capacity of all MoD vehicles to $C_v=3$ passengers. The GNL allocation weights are $\gamma_{jn}=0.5$ and
dissimilarity parameters $\mu_n=0.65$ for all nests. 
Utility coefficients are $\alpha_f=0.16~ \$^{-1}$, $\alpha_w=0.055~\mathrm{min}^{-1}$, $\alpha_v=0.035~\mathrm{min}^{-1}$, $\alpha_\delta=0.025~\mathrm{min}^{-1}$, $\alpha_t=0.100~\mathrm{min}^{-1}$, and $\alpha_\omega=0.080~\mathrm{min}^{-1}$~\cite{usdot2026bca}.
A minimum choice probability of $\underline{p}=0.01$ is imposed for each offered option.

\subsection{Impact of service portfolio design and pricing}
\begin{table}[tb]
    \centering
    \caption{Operator revenue, operating cost, and service margin under the four optimization methods. Values are mean $\pm$ standard deviation over five simulation runs.}
    \label{tab:profitability}
    \begin{tabular}{lrrr}
        \toprule
        \multirow{2}{*}{\textbf{Method}}
        & \textbf{Revenue $\uparrow$}
        & \textbf{Operating cost $\downarrow$}
        & \textbf{Service margin $\uparrow$} \\
        & \textbf{(\$1,000/h)}
        & \textbf{(\$1,000/h)}
        & \textbf{(\$1,000/h)} \\
        \midrule
        \textit{Standard fare}
        & $26.09 \pm 0.11$
        & $8.10 \pm 0.03$
        & $18.00 \pm 0.09$ \\
        \textit{Fare only}
        & $31.07 \pm 0.25$
        & $\mathbf{7.98 \pm 0.05}$
        & $23.10 \pm 0.22$ \\
        \textit{Menu only}
        & $27.65 \pm 0.23$
        & $8.13 \pm 0.05$
        & $19.52 \pm 0.20$ \\
        \cellcolor{gray!20} \textbf{\textit{Joint}}
        & \cellcolor{gray!20} $\mathbf{33.04 \pm 0.17}$
        & \cellcolor{gray!20} $\mathbf{7.98 \pm 0.04}$
        & \cellcolor{gray!20} $\mathbf{25.07 \pm 0.13}$ \\
        \bottomrule
    \end{tabular}
\end{table}
We compare four methods. \textit{Standard fare} offers all feasible options at standard fares, \textit{Fare only} optimizes fares while offering all feasible options, \textit{Menu only} optimizes the service menu while using standard fares, and \textit{Joint} jointly optimizes the service menu and fares. For each passenger $i$ and each service option $j$, the standard fare is calculated as $f_i^{j, \text{std}}(\phi) = b_j + r_j^D \ell_i^j (\phi) + r_j^T \tau_i^j(\phi)$, where $\ell_i^j(\phi)$ and $\tau_i^j(\phi)$ denote the total MoD travel distance and time of trip plan $\phi$, respectively~\cite{nyctlc_rates}. For $E$ and $ET$, we use $(b_j,r_j^D,r_j^T)=(\$3.00,\$2.20/\text{km},\$0.35/\text{min})$, while for $P$ and $PT$, we use $(\$2.00,\$1.45/\text{km},\$0.22/\text{min})$~\cite{nyctlc_rates, uber_pool_pricing}.

Table~\ref{tab:profitability} shows that jointly optimizing service menus and fares increases the service margin by $28.4\%$ over \textit{Menu only} and by $8.5\%$ over \textit{Fare only}. These gains are driven primarily by higher revenue rather than lower operating costs. Compared with \textit{Menu only}, \textit{Joint} increases revenue by $19.5\%$ with only a small reduction in operating cost. Compared with \textit{Fare only}, revenue increases by $6.3\%$, while operating cost remains nearly unchanged. A supplementary video of \textit{Joint} operations is available at \href{https://youtube.com/shorts/gSySK2ZPlO4?feature=share}{\textcolor{blue}{this link}}.

\subsection{Impact of passenger choice model}
\begin{figure}[tb]
    \centering
    \begin{subfigure}[t]{0.49\linewidth}
        \centering
        \includegraphics[width=\linewidth]{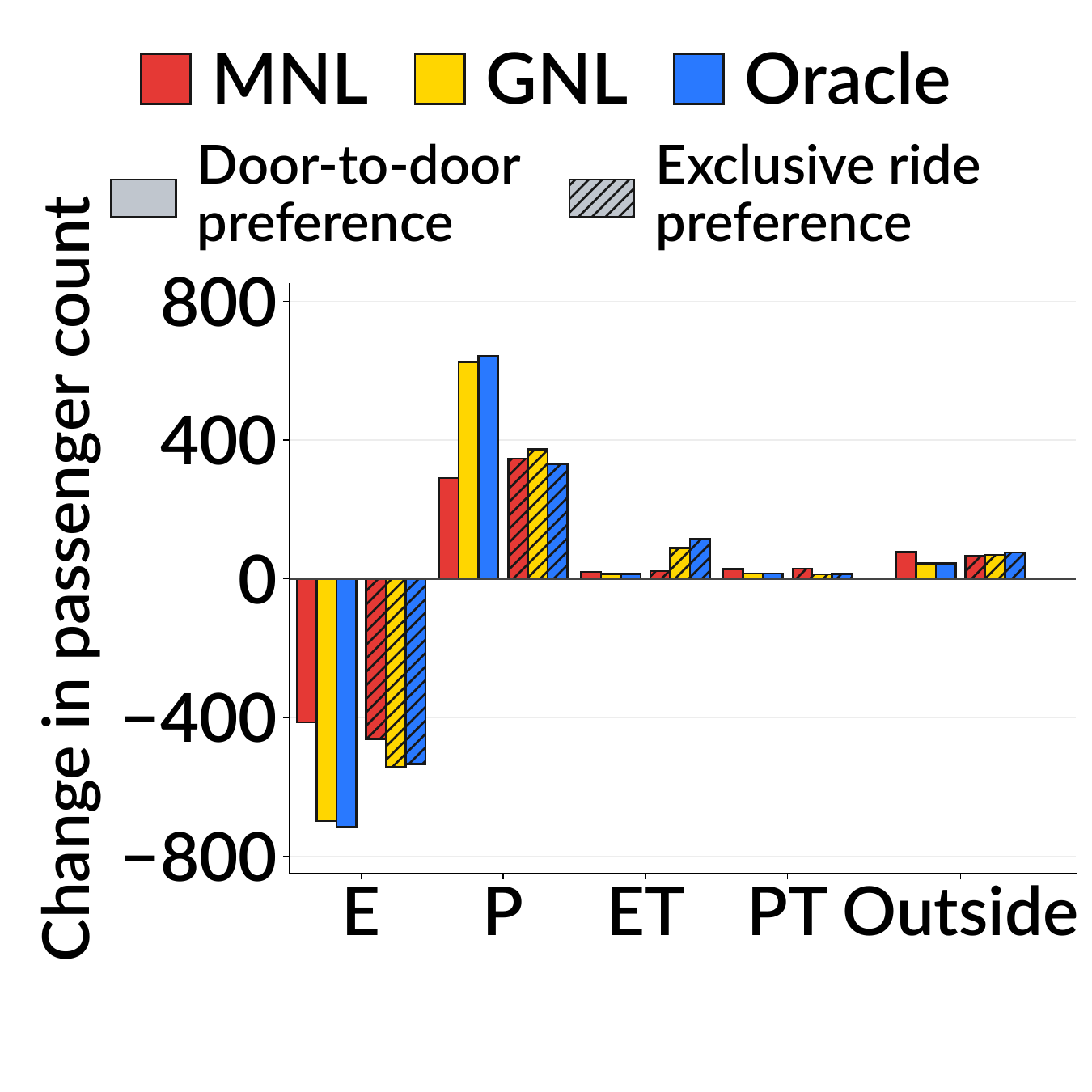}
        \caption{Changes in expected passenger counts under MNL, GNL, and \textit{Oracle}}
        \label{fig:demand_shift}
    \end{subfigure}
    \begin{subfigure}[t]{0.49\linewidth}
        \centering
        \includegraphics[width=\linewidth]{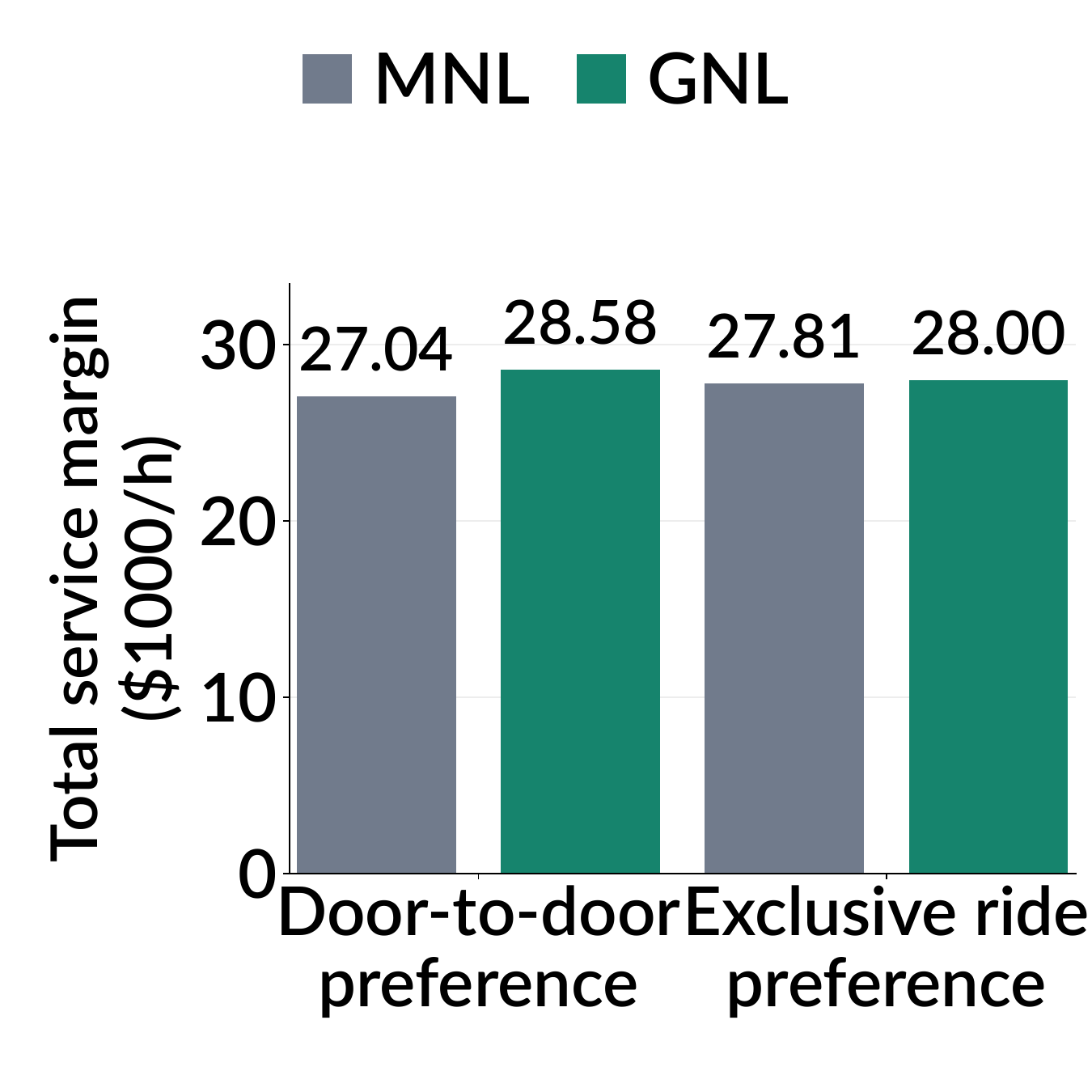}
        \caption{Service margin under MNL and GNL}
        \label{fig:gnl_service_margin}
    \end{subfigure}
    \caption{Passenger responses and operator service margins following a $50\%$ increase in the fare of $E$ under door-to-door and exclusive preference scenarios. Choice changes are reported as counts over the $4{,}900$ requests.}
\end{figure}
To examine the importance of modeling correlated substitution patterns, we evaluate how different choice models predict passenger responses to a fare change. Specifically, we increase the optimized fare of the exclusive door-to-door option ($E$) by $50\%$ and examine how demand is redistributed across the alternatives. We consider two scenarios by varying the GNL allocation weights and dissimilarity parameters. In the door-to-door preference scenario, we set $(\gamma_{jn},\mu_n)=(0.9,0.3)$ for the door-to-door and transit-connected nests and $(0.1,1.0)$ for the exclusive and pooled nests, strengthening substitution along the transit-use dimension. In the exclusive ride preference scenario, these settings are reversed, strengthening substitution along the vehicle-sharing dimension. Passenger choices are generated using a GNL model with parameters specified for each scenario, referred to as \textit{Oracle}. The GNL model used by the operator instead estimates these parameters from $2{,}000$ simulated choice observations, while the MNL model assumes that demand shifts proportionally across the remaining alternatives, regardless of their similarities.

Fig.~\ref{fig:demand_shift} shows the resulting changes in expected passenger choices. Under the door-to-door preference scenario, passengers leaving $E$ shift predominantly toward the pooled door-to-door option ($P$). The estimated GNL closely reproduces this substitution pattern, whereas MNL predicts a substantially smaller shift toward $P$. Under the exclusive ride preference scenario, the distinction is most evident for the exclusive transit-connected option ($ET$). GNL and \textit{Oracle} predict a considerably larger increase in $ET$ demand than MNL. Thus, GNL better captures how substitution patterns change with passenger preferences, whereas MNL remains constrained by proportional substitution.
Fig.~\ref{fig:gnl_service_margin} shows that the profitability benefit of improved choice prediction differs across the two scenarios. In the door-to-door preference scenario, GNL more accurately captures the shift from $E$ to $P$, allowing the operator to set fares that better reflect this substitution pattern. As a result, GNL achieves about $\$1,500$ higher service margin per hour than MNL. In the exclusive ride preference scenario, GNL improves the prediction of the shift from $E$ to $ET$, but the service-margin gain is only about $\$200$ per hour. This smaller improvement arises because $ET$ provides only part of the trip through MoD, while the transit fare is paid to the transit agency rather than retained by the MoD operator. Thus, even though GNL predicts the shift toward $ET$ more accurately than MNL, the operator has limited opportunity to increase its service margin through fare optimization. These results show that the profitability gains from improved choice prediction depend on both passenger substitution patterns and the revenue retained by the operator across services.

\subsection{Computational efficiency of the pricing algorithm}
\begin{figure}[t]
    \centering
    \begin{subfigure}[t]{0.49\linewidth}
        \centering
        \includegraphics[width=\linewidth]{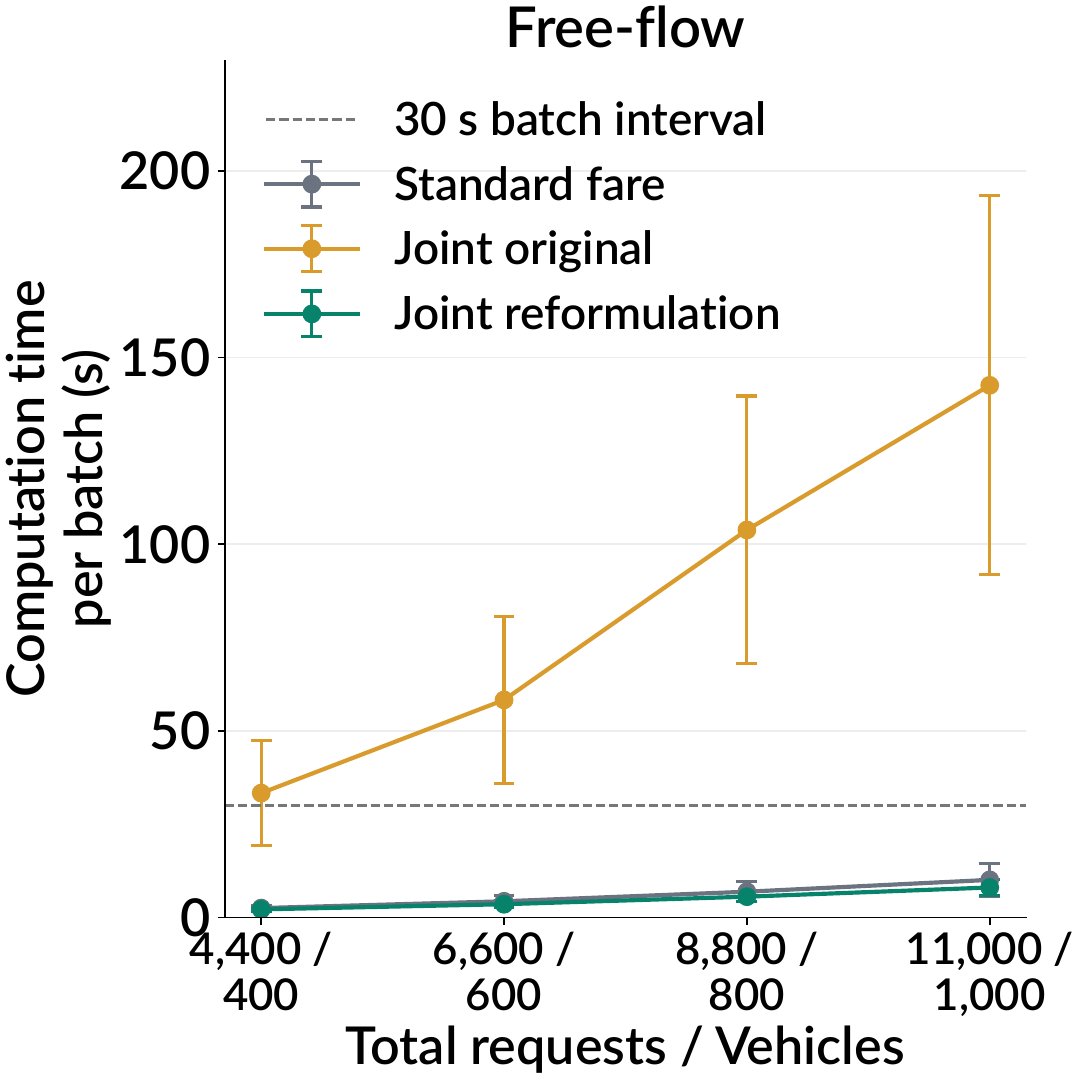}
        \caption{Average computation time under free-flow conditions.}
        \label{fig:imp_joint_service_margin}
    \end{subfigure}
    \begin{subfigure}[t]{0.49\linewidth}
        \centering
        \includegraphics[width=\linewidth]{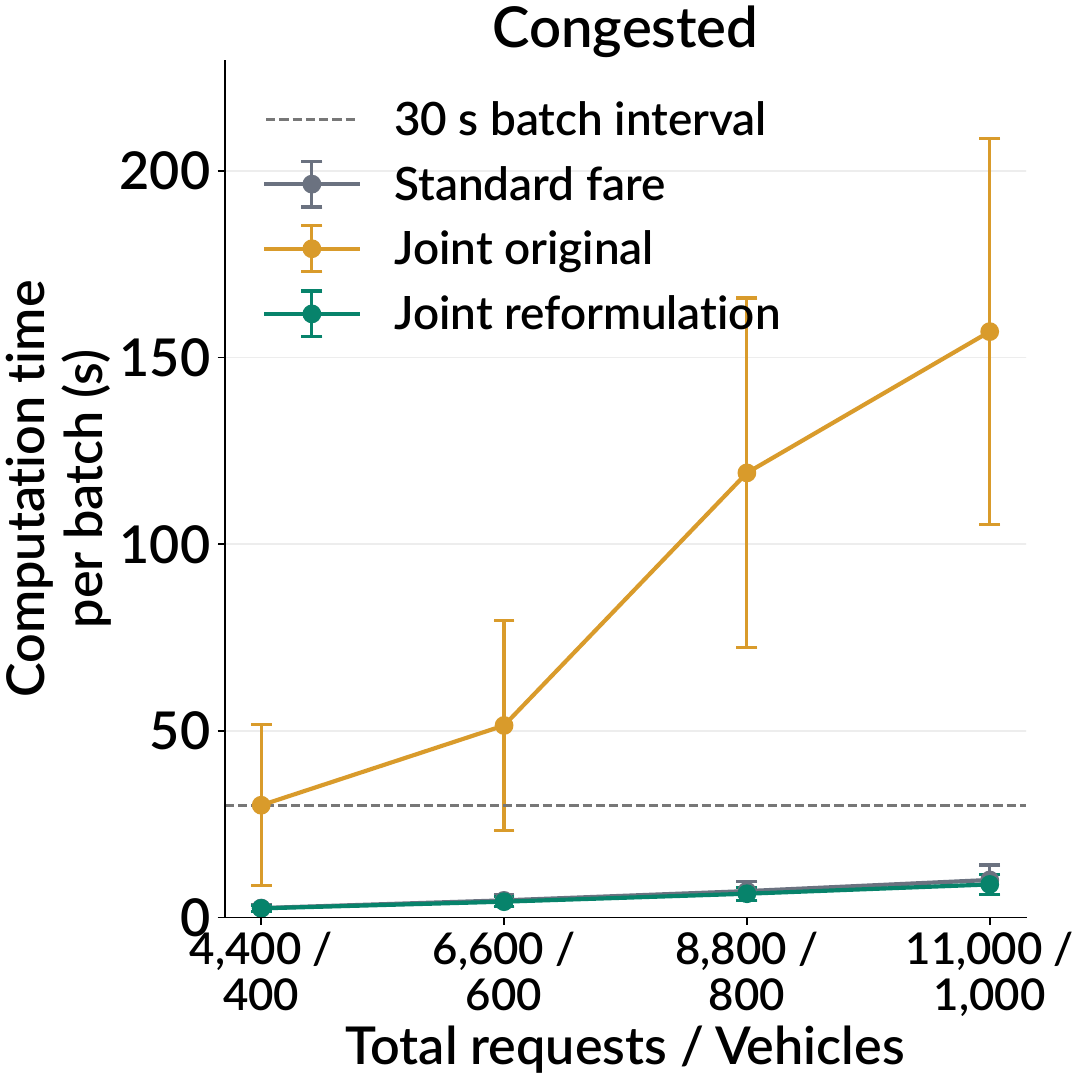}
        \caption{Average computation time under congested conditions.}
        \label{fig:imp_joint_rev_cost}
    \end{subfigure}
    \caption{Average computation time per batch for the original fare-space optimization (\textit{Joint original}) and the generalized-entropy reformulation (\textit{Joint reformulation}) across four demand levels at a fixed request-to-vehicle ratio of $11$:$1$. The dashed line marks the $30$~s batch interval, which sets the time limit for solving each batch optimization problem.}
    \label{fig:comp_eff}
\end{figure}
Fig.~\ref{fig:comp_eff} shows that the generalized-entropy reformulation (\textit{Joint reformulation}) reduces the average runtime per batch compared with the original fare-space optimization (\textit{Joint original}) under both free-flow and congested conditions. The proposed reformulation achieves a $12.2$--$18.8$ times speedup across all tested demand levels while producing numerically identical objective values. Under both traffic conditions, \textit{Joint original} exceeds the $30$ s batch interval even at $4{,}400$ requests, whereas \textit{Joint reformulation} remains within the interval across all tested demand levels. As demand increases, both formulations require more computation as the optimization problem grows. However, the original formulation directly optimizes fares together with fare-dependent choice probabilities, resulting in substantially longer runtimes. The reformulation removes products of fares and fare-dependent choice probabilities from the objective by optimizing in the choice-probability space, improving computational tractability for real-time operation.

% Conclusions
\section{CONCLUSIONS}
This paper proposed a real-time framework for jointly optimizing service menus, fares, and vehicle trip plans in multimodal MoD systems while accounting for correlated passenger substitution. The results show that jointly considering service portfolio design and pricing improves operator profitability, and that the generalized-entropy reformulation substantially reduces computation time while preserving mathematical equivalence to the original fare-space formulation.
Future work will focus on two main extensions. First, collaboration with real-world mobility service providers could enable the collection of observed passenger choice data to estimate and validate the choice model under actual service conditions. Second, the framework can be extended to account for how current decisions affect future vehicle availability and demand, enabling more forward-looking operational decisions.
% Appendix
% \input{sections/appendix}

% Acknowledgement
% \input{sections/acknowledgement}

\bibliographystyle{IEEEtran}
\bibliography{IEEEabrv, ref}

\end{document}